\documentclass{PoS}

\pdfoutput=1

\usepackage[utf8]{inputenc}

\usepackage{amsmath,amssymb}
\usepackage{units}
\usepackage{cleveref}

\newcommand\emailUsername[2]{{\tt\href{mailto:#1@#2}{#1}}}
\newcommand\emailFullpath[2]{{\tt\href{mailto:#1@#2}{#1@#2}}}

\newcommand{\tcl}{$1.5 \; T_c$}

\newcommand{\tch}{$3.0 \; T_c$}

\newcommand{\MSb}{{\overline{MS}}}
\newcommand{\GeV}[1]{{\unit[#1]{GeV}}}
\newcommand{\MeV}[1]{{\unit[#1]{MeV}}}

\DeclareMathOperator{\ofOrder}{O}

\title{The thermodynamic and the continuum limit of meson screening masses}

\ShortTitle{The thermodynamic and the continuum limit of meson screening masses}

\author{O. Kaczmarek, E. Laermann, \speaker{M. M\"uller}\\
Fakult\"at f\"ur Physik, Universit\"at Bielefeld, D-33615 Bielefeld, Germany\\
E-mail: 
\emailUsername{okacz}{physik.uni-bielefeld.de},
\emailUsername{edwin}{physik.uni-bielefeld.de},
\emailFullpath{mmueller}{physik.uni-bielefeld.de}
}

\abstract{
We present results on the thermodynamic and continuum limit 
of meson screening masses in the deconfined phase,
using standard staggered and non-perturbatively
clover-improved Wilson fermions in the quenched
approximation with light quark masses.

For two temperatures, 1.5 Tc and 3.0 Tc, it is found
that on finite lattices screening masses differ between
the actions. We study if both actions reproduce 
the same masses in the continuum by employing 
different methods of extrapolation
to the thermodynamic and continuum limit.
}

\FullConference{31st International Symposium on Lattice Field Theory - LATTICE 2013\\
		July 29 - August 3, 2013\\
		Mainz, Germany}

\begin{document}

\section{Introduction}

Meson screening masses, since long, have been calculated in finite
temperature QCD, being a measure of hadronic quark-gluon plasma
excitations. The methods employed reached from
hard thermal loop calculations \cite{Alberico:2007yj} to approaches by
dimensional reduction \cite{Vepsalainen:2007ke} to lattice simulations,
see e.g. \cite{Cheng:2010fe,Gupta:2013vha} for the latest studies with
dynamical staggered quarks.

In finite temperature lattice calculations
meson screening masses can be determined with high precision. 
However, it was observed that
on finite lattices the results obtained with different discretization
schemes, the Wilson action 
with non-perturbative clover improvement and
the standard staggered action in particular, did not agree.
These deviations are assumed to be caused by different
discretization effects.
Although both actions share that to leading order the finite lattice
spacing effect is $\ofOrder(a^2)$, the pre-factors
may differ. Differences can also emerge due to
varying finite volume effects.

Here we present a systematic study in quenched lattice QCD to investigate
this assumption by carrying out 
the thermodynamic as well as the continuum limit
for both actions.

\section{Lattice setup}

At two temperatures in the deconfined phase, \tcl{} 
and \tch{}, meson screening masses have been calculated
for the pseudoscalar (PS), scalar (S), transverse vector
(V) and transverse axialvector (AV) channel. 

For the low temperature, a set of five 
aspect ratios $N_\sigma/N_\tau= 2,3,4,6,8$ 
allowed to take the thermodynamic limit, 
while four temporal lattice extents $N_\tau = 8,10,12,16$ 
were selected for the continuum extrapolation. 
For the high temperature, a reduced set 
of four aspect ratios $N_\sigma / N_\tau = 2,3,4,8$
and three temporal extents $N_\tau = 8,12,16$
proved to be sufficient for the analysis. Ensembles
of 100 to 350 gauge field configurations have been analyzed at each
lattice size.

For both fermion actions, the non-perturbatively improved
clover action \cite{Luscher:1996ug} and the standard staggered discretization,
the investigation was carried out for 
light (valence) quark masses,
$m_{q,\MSb}(\mu=\GeV{2}) < \MeV{25}$. 
At and below this value no noticeable quark mass effect on the screening
masses was found.


\section{Meson screening masses}

The meson screening masses are extracted from meson correlators,
evaluated along one of the spatial axes (chosen as $n_z$ in the
following) and projected to zero `momentum' ($p_x = p_y = p_\tau = 0$).
On a finite lattice with periodic boundary conditions the correlator for
Wilson-type fermions
reads
\begin{equation}
G^{\rm Wilson}(n_z) 
= \sum_i A_i \left( e^{-m_i n_z} + e^{-m_i (N_\sigma-n_z)}\right)
= \sum_i A^\prime_i \cosh( (N_\sigma/2 - n_z) \cdot m_i).
\label{eq-meson-masses}
\end{equation}
The steepness of the exponential fall-off is
given by the (screening) masses $m_i$ of the different states.
The ground state $m_0$ dominates the correlator for
large separations $n_z$.
For fermions of staggered type the situation is slightly more complicated
due to the presence of socalled parity partners that contribute with an
oscillating part $\sim (-1)^{n_z}$ to the correlator,
\begin{align}
G^\text{stagg.}(n_z) 
= \sum_i \Big ( & A^{\prime,\text{no}}_i \cosh( (N_\sigma/2 - n_z) \cdot m^\text{no}_i )  + 
(-1)^{n_z} A^{\prime,\text{os}}_i \cosh( (N_\sigma/2 - n_z) \cdot m^\text{os}_i ) \Big )
\label{eq-meson-masses-stagg}
\end{align}

A convenient way to extract the screening masses is to fit the
$A^\prime_i$, $m_i$ in 
\cref{eq-meson-masses,eq-meson-masses-stagg}
to a correlator measured on the lattice. 
These fits can be performed by taking a varying number of
states into account. In the following, one-state
fits refer to taking into account only the ground state contribution(s) $m_0$, and
two-state fits refer to taking into account also one excitation. For two-state
staggered fits, only the dominant, either oscillating or non-oscillating part is fitted with
two states contributing.


\section{One-state fits}

\begin{figure}
\includegraphics{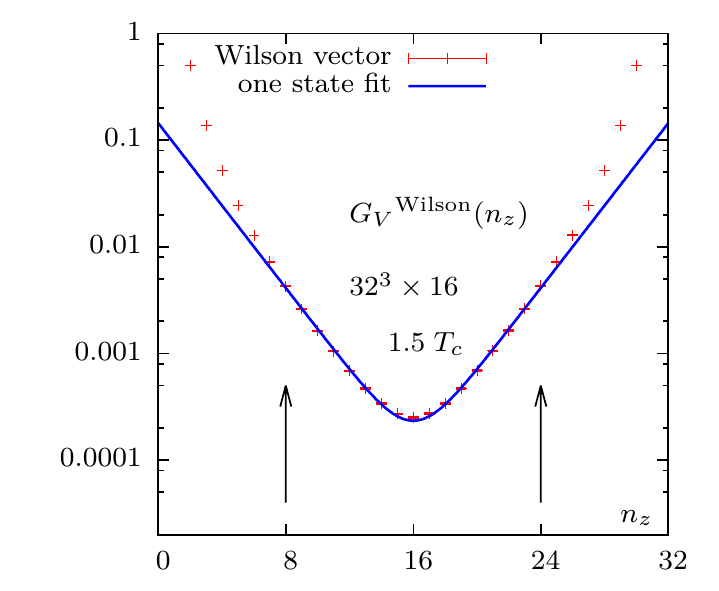}
\includegraphics{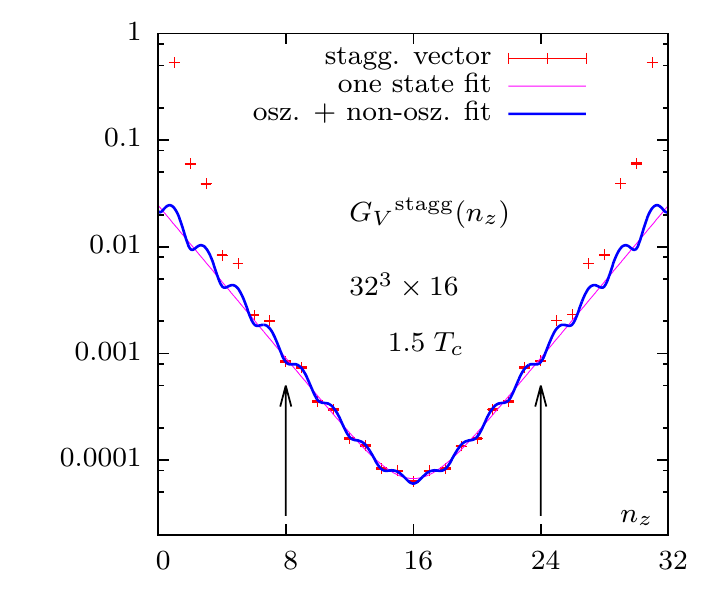}
\caption[One-state fit of the V correlator, \tcl{}, $32^3\times 16$]{A fit 
to the vector correlation function measured
on a $32^3\times 16$ lattice at \tcl{}. Wilson correlator (left) fitted
with \cref{eq-meson-masses},
and staggered correlator (right) with an additional oscillating ground
state contribution through \cref{eq-meson-masses-stagg}. The arrows mark the
fit window $\frac{1}{4}N_\sigma\dots \frac{3}{4}N_\sigma$.}
\label{fig-oneStateFits}
\end{figure}

Fitting \cref{eq-meson-masses,eq-meson-masses-stagg} to data,
care has to be taken in selecting an appropriate 
fit window $n^\prime_z \; \dots  \; N_\sigma-n^\prime_z$. 
Varying the fit window can be used to find a plateau, 
where the mass will level off indicating that
the fit has stabilized to a ground state, see \cref{fig-fitWindowTwoStates}.
However, on smaller lattices no such plateau is found,
since the necessary separation $n^\prime_z$ from the source can not be
chosen large enough.

Thus, a different approach -- inspired by the analysis of
effective masses in the high temperature, free limit -- is used: 
A fixed, consistent value
of $n^\prime_z=N_\sigma/4$ was chosen for all lattices,
motivated by ground state plateau onset on larger lattices.
On lattices with smaller aspect ratios, these fits
will inevitable  pick up contributions from exited states
and merge them into a higher mass $m_0$.
The contributions from these higher states
are removed in the next step of the analysis, by extrapolating 
towards the thermodynamic limit.


\section{Thermodynamic limit}

\begin{figure}
\includegraphics{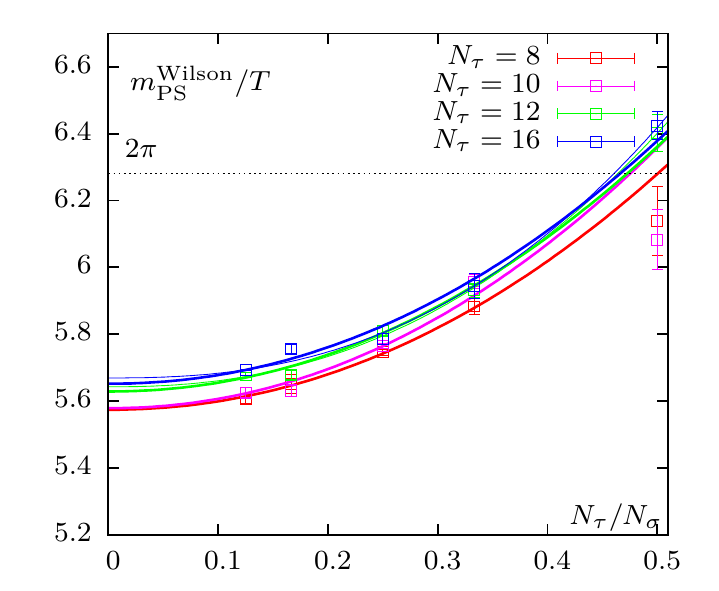}
\includegraphics{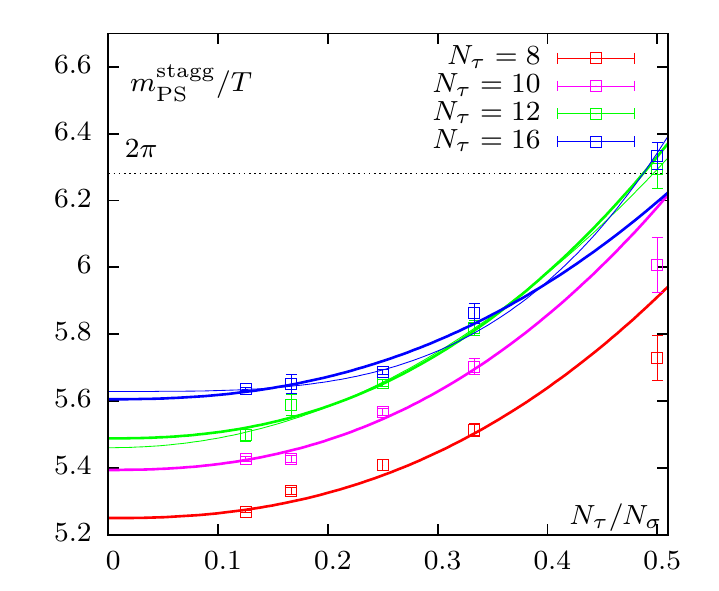}
\caption[Thermodynamic limit of the PS,
at \tcl{}]{Thermodynamic limit obtained by fitting \cref{eq-ansatz-thermLimit} to
Wilson (left) and staggered (right) pseudoscalar meson screening masses at \tcl{}. 
The thick lines are obtained from a full
fit with one shared exponent $c$ for all lattice spacings. 
The thin lines are obtained by
fits with a dataset limited to $N_\tau=12$ and $N_\tau=16$ to crosscheck the
assumption that the exponent $c$ is independent of the lattice spacing.
The dotted line at $2\pi$ marks the free theory limit.
}
\label{fig-thermLimit-oneState}
\end{figure}

An ansatz for the thermodynamic limit $N_\sigma \rightarrow \infty$ 
now has to be found and motivated: 
For the free theory, 
the leading dependence on $N_\sigma$ for the screening masses
extracted as explained above
is linear in $1 / N_\sigma$. 
From zero temperature simulations, 
the mass at finite volume is found to scale 
with the third power of the spatial lattice extent
\cite{Fukugita:1992jj}.
For finite temperature, it is therefore sensible to choose
\begin{equation}
m_{N_\sigma/N_\tau} = m_{N_\sigma \rightarrow \infty / N_\tau} 
\left ( 1 + b_{N_\tau} \cdot (N_\tau / N_\sigma)^c \right )
\label{eq-ansatz-thermLimit}
\end{equation}
as an ansatz for the fit, with three free 
parameters $m_{N_\sigma \rightarrow \infty / N_\tau}$, $b_{N_\tau}$ and $c$.

The thermodynamic limit $m_{N_\sigma \rightarrow \infty / N_\tau}$
is still subject to finite spacing effects, therefore a separate
value has to be obtained for every temporal lattice 
extend $N_\tau = 8, 12, \dots$. However, a combined fit
can be performed to reduce the total number of free parameters
and stabilize the fit: As indicated by the index, 
$b_{N_\tau}$ in \cref{eq-ansatz-thermLimit} is allowed to vary for
different $N_\tau$. The exponent $c$, on the other hand,
should only depend on the temperature and thus is shared
among all $N_\tau$.

In \cref{fig-thermLimit-oneState}
the result of these fits can be found for the pseudoscalar at \tcl{}. 
The figure also shows (as thin lines) two
fits in which the exponent $c$ was determined with the 5 data 
points from $N_\tau=12$
and $N_\tau=16$ lattices only, in order to crosscheck whether the 
procedure of fixing the exponent
to be independent of the lattice spacing is correct. 


\section{Two-state fits}

\begin{figure}
\includegraphics{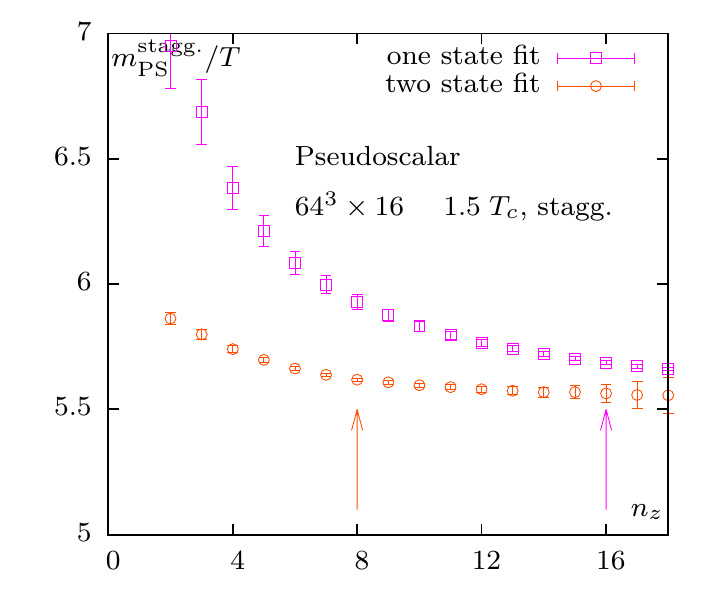} 
\includegraphics{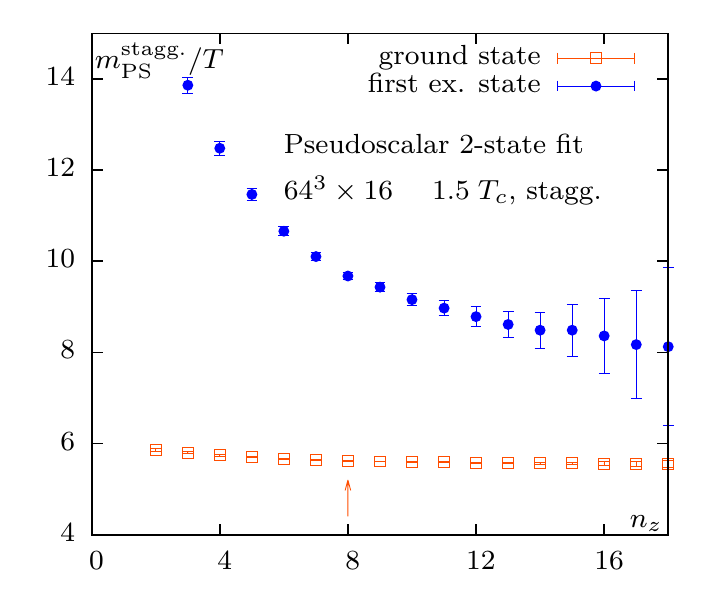}
\caption[Fit range dependence of PS screening masses, \tcl{}, $64^3\times 16$]{
Staggered pseudo-scalar screening masses, 
obtained at \tcl{} on $64^3\times 16$ lattices,
for different sizes $n_z \dots N_\sigma-n_z$ of the fit window.
\textbf{Left:} The ground state results of the 
one- and two-state fits are compared, where the one-state
fits show a much stronger dependence on the fit window size. 
\textbf{Right:} The ground state and the exited state
of the two-state fits are compared, showing that 
when increasing the fit window size, higher
contributions mainly enter into the existed state,
while the ground state is mostly uneffected.
The arrows mark the fit window position $n_z$ chosen
for the analysis.
}
\label{fig-fitWindowTwoStates}
\end{figure}

A second approach to extract the ground state 
is to fit 
the correlator 
to
\cref{eq-meson-masses,eq-meson-masses-stagg} 
including one excitation.
In \cref{fig-fitWindowTwoStates}, different 
fit window sizes have been tested
for both one-state and
two-state fits. As expected,
the mass found by one-state fits 
is clearly influenced by the fit window
size, as it raises towards the source. 
For the two-state fits,
the exited state $m_1$  is found to absorb the higher
contributions, while the ground state is 
stable over a wide region of
fit window sizes. 
This motivates to set 
$\frac{1}{2}N_\tau \; \dots \; N_\sigma-\frac{1}{2}N_\tau$
as a fit window for two-state fits, to 
ensure a constant physical distance from the source
for all aspect ratios and lattice spacings.

The procedure fails for very 
small aspect ratios ($N_\sigma/N_\tau=2,3$), 
where the system is too small to
show a clear ground state.
A minimal aspect ratio of $N_\sigma/N_\tau=4$
seems to suffice for larger $N_\tau=12,16$, while
$N_\sigma/N_\tau=6,8$ give more precise results, especially
on coarse lattices $N_\tau=8,10$.

For these larger aspect ratios,
the fit results show no clear dependence on the
lattice volume, thus they are averaged.
For the pseudo-scalar and scalar channels as well as all 
staggered channels, results from
the two-state fits are compatible with the thermodynamic
limit of one-state fits shown in \cref{fig-thermLimit-oneState}.
For the Wilson vector and axial-vector channel, results from
the one-state fits are somewhat lower then from the two-state
fits. Looking ahead on the final comparison of both actions,
the thermodynamic limit through \cref{eq-ansatz-thermLimit}
probably overestimated the finite volume effects.

\section{Continuum limits}

After the thermodynamic limit or the ground state mass from a two-state
fit have been obtained, they have to be extrapolated to the continuum.

\begin{figure}
\includegraphics{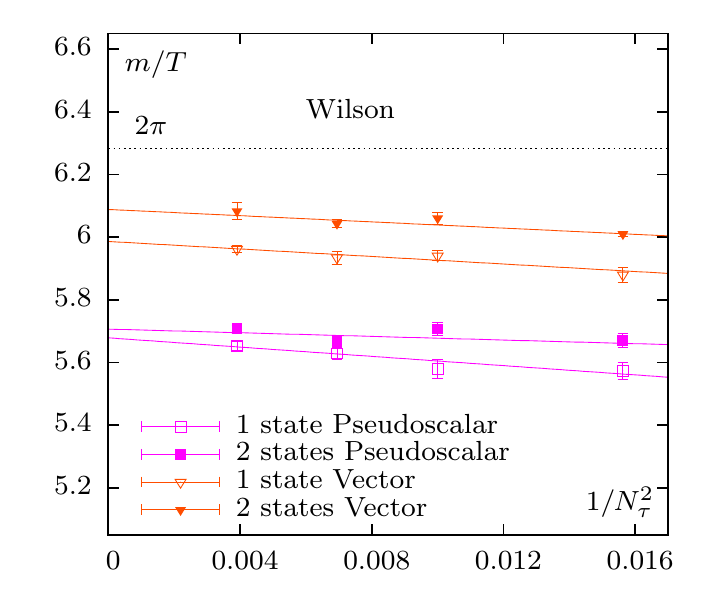}
\includegraphics{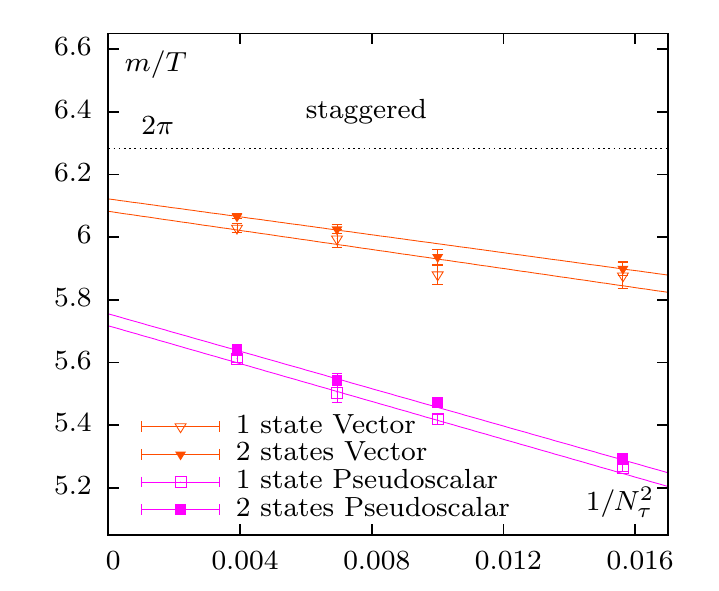}
\caption{
Continuum extrapolation using \cref{eq-cont-extr} of
the pseudoscalar and vector channels,
for both Wilson (left) and staggered (right) fermions, at \tcl{}. 
The data points are the result of the thermodynamic limit
(infinite volume extrapolation of the one-state fits) and the ground
state masses obtained from two-state fits.
The dotted line marks the free theory mass of $2\pi T$.
}
\label{fig-cont-limit}
\end{figure}


Both the clover-improved Wilson action as well as the standard
staggered fermion action show $\ofOrder(a^2)$
discretization errors.
At a fixed temperature, since $T = 1 / (a \cdot N_\tau )$, the continuum
is reached by extrapolating in $1/N_{\tau}^2$, so
\begin{equation}
m_{N_\tau} = m^\text{cont.} \cdot \big( 1 + d \cdot \frac{1}{N_{\tau}^2} \big )
\label{eq-cont-extr}
\end{equation}

Higher orders as $\ofOrder(a^4)$ 
or $\ofOrder(a^2 \ln(a))$  
could enter into \cref{eq-cont-extr}, but these contributions
seem to be small enough to not influence the results.

The extrapolation results are shown in \cref{fig-cont-limit}.
Comparing Wilson and staggered fermions in these plots, it becomes very apparent that
the staggered action is much more strongly dependent on the lattice spacing, while
Wilson fermions seem much less affected. 

\section{Results and conclusions}

\begin{figure}[t]
\includegraphics{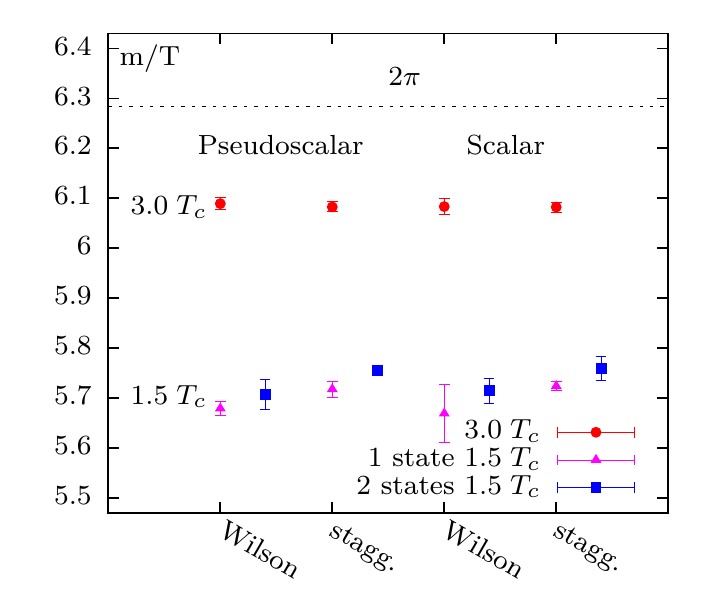}
\includegraphics{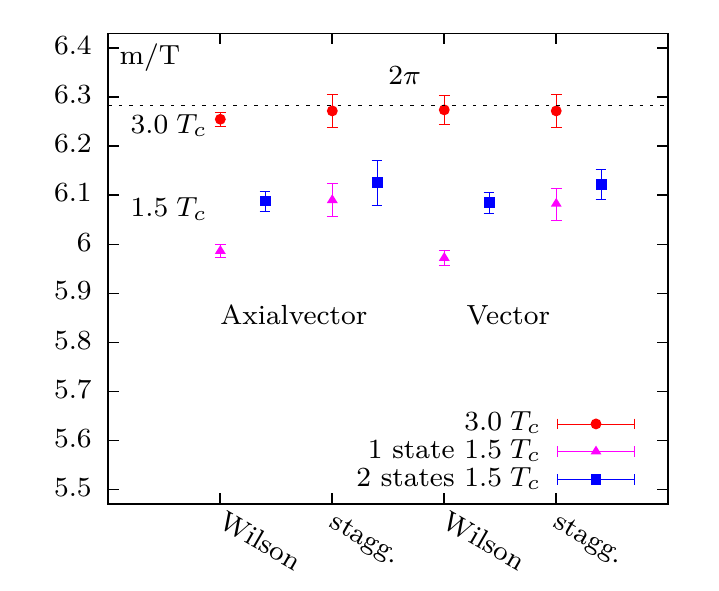}
\caption[Meson screening masses final dataset]{
Continuum limit for the (pseudo)scalar and (axial)vector
for both the Wilson and the staggered action. At \tcl{},
both the masses from the thermodynamic limit (1 state fits)
and the ground state mass from two-state fits have been extrapolated
to the continuum. At \tch{}, masses have only been extrapolated
from the thermodynamic limit.
}
\label{fig-screening-final}
\end{figure}

The final results of this study are meson
screening masses extrapolated to the thermodynamic and
continuum limit, for the pseudo-scalar, scalar, vector and
axial-vector channel, for both the Wilson and the
staggered action, and at both temperatures \tcl{} and \tch{}.
In \cref{fig-screening-final}  all final data points 
are summarized. 

The differences between both actions on finite lattices vanish
when the ground state masses are extracted (by taking the 
thermodynamic limit or performing a two-state fit) and the 
continuum limits are carried out. 
The staggered action shows much more pronounced lattice spacing effects,
so the continuum limit is a crucial step. Finite lattice volume effects
are similar for both actions.

The thermodynamic limit can be reached by extrapolation through
ansatz \cref{eq-ansatz-thermLimit} or by fitting multiple states
to a correlator on a reasonably sized lattice. 
Except for the Wilson (axial)vector, these two-state fits reach 
results compatible
to the full thermodynamic limit, when carried out
at aspect ratios of $N_\sigma / N_\tau \geq 4$
and at temporal extends $N_\tau \geq 12$.
If this holds for other temperatures and actions, it might help to reduce the
lattice sizes and thereby computing cost needed in screening masses studies. 

Both the pseudoscalar and the scalar
masses as well as the vector and the axialvector mass are clearly 
degenerate at both temperatures \tcl{} and \tch{}. This
holds not only for the continuum extrapolation, but already for finite
size lattices.


HTL and DR results \cite{Alberico:2007yj,Vepsalainen:2007ke} 
currently predict 
the free theory meson screening masses $2 \pi T$ to be reached from above,
for both the (pseudo)scalar and the (axial)vector. At \tcl{}, none of the masses reach
this free theory limit, they are both clearly below.
For \tch{}, all masses are closer to $2 \pi T$ and the vector and axial-vector are
compatible with $2 \pi T$. 

The free theory also predicts
a full degeneracy between scalar, vector, pseudoscalar and axialvector,
so a common mass for all four channels should be seen when approaching
the free limit. This is not observed at either temperature. 

\section{Acknowlegdements}

The results 
have been 
achieved by using JUGENE resources at the 
J\"ulich Supercomputing Centre 
and GPU-cluster resources of the 
lattice gauge theory group at Bielefeld University.
This work is supported by the 
IRTG/GRK 881 "Quantum Fields and Strongly Interacting Matter".

\bibliography{bibScreeningMasses}
\bibliographystyle{JHEPmod}

\end{document}